\documentclass{elsart}

\usepackage{graphicx}
\usepackage{epsfig,psfig}
\usepackage{amssymb}
\newcommand{\be}{\begin{equation}}
\newcommand{\ee}{  \end{equation}}
\newcommand{\ba}{\begin{eqnarray}}
\newcommand{\ea}{  \end{eqnarray}}
\newcommand{\bc}{\begin{center}}
\newcommand{\ec}{  \end{center}}

\journal{Annals of Physics (N. Y.)}

\begin{document}
\begin{frontmatter}
\title{Random--Matrix Approach \\to RPA equations. I}
\author[a]{X. Barillier--Pertuisel}, 
\author[b]{O. Bohigas}, 
\author[c]{H. A. Weidenm{\"u}ller}

\address[a]{Institut de Physique Nucl\'{e}aire, IN2P3-CNRS, 
UMR8608\\Universit\'{e} Paris-Sud,F-91406 Orsay, France}
\address[b]{Laboratoire de Physique Th\'eorique et Mod\`eles Statistiques, 
UMR8626\\Universit\'{e} Paris-Sud,F-91405 Orsay, France}
\address[c]{Max--Planck--Institut f{\"u}r Kernphysik, Heidelberg, Germany}

\begin{abstract}
We study the RPA equations in their most general form by taking the
matrix elements appearing in the RPA equations as random.  This yields either a 
unitary or an orthogonally invariant random--matrix model that does not appear 
in the Altland--Zirnbauer classification. The average spectrum of the model is
studied with the help of a generalized Pastur equation. Two
independent parameters govern the behaviour of the system: The
strength $\alpha^2$ of the coupling between positive-- and
negative--energy states and the distance between the origin and the
centers of the two semicircles that describe the average spectrum for
$\alpha^2 = 0$, the latter measured in units of the equal radii of the
two semicircles. With increasing $\alpha^2$, positive-- and
negative--energy states become mixed and ever more of the spectral
strength of the positive--energy states is transferred to those at
negative energy, and vice versa. The two semicircles are deformed and
pulled toward each other. As they begin to overlap, the RPA equations
yield non--real eigenvalues: The system becomes unstable. We determine
analytically the critical value of the strength for the instability to
occur. Several features of the model are illustrated numerically.
\end{abstract}

\begin{keyword}
Random--matrix theory; Random Phase Approximation
\PACS 21.10.-k, 21.60.-n
\end{keyword}
\end{frontmatter}

\section{Introduction}

Random--matrix theory (RMT), introduced into physics by Wigner in the
1950s, has found an ever increasing range of applications. In
particular, it has been widely used to describe spectral fluctuation
properties of many--body systems (for a review see Ref.~\cite{Guh}).
The main ingredients of RMT are general symmetry properties and the
assumption that no state in Hilbert space plays a preferred role. The
success of this approach can be ascribed to the fact that many
physical systems are rather featureless and to a large extent
chaotic~\cite{BGS}. Most theoretical work in many--body systems has,
on the other hand, been devoted to non--statistical aspects, in
particular to identifying and describing collective features. A
prominent example is that of the giant collective states in atomic
nuclei. An extreme treatment is that of the so--called schematic
model~\cite{Bro}: A special (separable) form of the (residual)
interaction is responsible for the collective features and a
particular (collective) state plays a distinct role. A widely used
many--body approximation scheme used in this context is the
Random--Phase Approximation (RPA) first introduced to describe the
collective plasma oscillations of an electron gas. The RPA describes
the response of the system to a time--dependent field and can be
derived, for instance, with the help of a quasi--Boson approximation
within the equation--of--motion method~\cite{Row} or the time--dependent
Hartree--Fock equations ~\cite{Boh}.

This paper is the first of a series of two that are written in an
attempt to bridge the gap between the two extremes of a purely
statistical (and, thus, ``democratic'') and a highly special dynamical
description (the interesting features reside in a single state). We
define a random--matrix approach to the RPA equations by taking the
residual interaction matrix elements appearing in the RPA as random.
This may look like an oxymoron: The RPA is designed to describe
collective features whereas RMT is fundamentally ``democratic''.
However, in the second paper of this sequel (Ref.~\cite{BPX}, in
preparation) collective features will be implemented in our model, and
the present study of the RMT approach to RPA equations is a
prerequisite for that work. Our paper is not the first to address the
interplay between collective features and the background of ``chaotic
states'', theoretically as well as experimentally (see, for instance,
Refs.~\cite{Tou} for Gamow--Teller transitions, Ref.~\cite{Ric} for
giant resonances, Ref.~\cite{Har} for analogue states and
Ref.~\cite{Pac} and the literature therein on doorway states) and must
be seen in this context. To the best of our knowledge, however, this
paper is the first where the problem is addressed within a purely
random version of RPA.

Our random--matrix model possesses the symmetry of the RPA equations.
We study the average spectrum of the model in the limit of large
matrix dimension with the help of a generalized Pastur
equation ~\cite{Past} and compare the results with numerical
simulations. We are especially interested in the instability of the
RPA equations. That instability occurs when one eigenvalue becomes
imaginary as we contiuously vary a coupling parameter. We derive a
general criterion for the instability to develop. Changes of the
coupling parameter cause the system to go through four different
``phases'': i) the entire spectrum is on the real axis, ii) one part
is on the real axis and the other on the imaginary axis, iii) one part
is on the real axis, another on the imaginary axis and a third one in
the complex plane, and iv) the entire spectrum is on the imaginary
axis. We are not aware of other systems, or models, showing this rich
behaviour.

The matrix governing the RPA equations is non--Hermitean and 
so is, therefore, our random--matrix model. A similar model (but with 
restrictions that do not allow the eigenvalues to become imaginary) was studied 
in Ref.~\cite{Lue06}. Our model does not belong to the ten canonical 
Altland--Zirnbauer ensembles (see Ref.~\cite{Hei05}). We believe that, quite 
aside from the RPA equations, our work is of interest as a case study in 
non--Hermitean random--matrix models.

To make the paper self--contained and to define the notation we start
out with a brief resume of the derivation of the RPA equations in
Section~\ref{rpa}. In Section~\ref{ex} we review the symmetries of
these equations and draw conclusions about the distribution of
eigenvalues. In Section~\ref{rma} the random--matrix approach to RPA
is defined. In Section~\ref{pas} we derive the equations for the
spectral density using the average Green's function (Pastur
equations). In Sections~\ref{symm} and \ref{prop}
properties of the solutions of the Pastur equations are discussed.
These are illustrated in Section~\ref{num} with numerical simulations.
In Section~\ref{crit} we establish the critical value of the coupling
strength at which instabilities occur. Summary and conclusions are
given in Section~\ref{summ}. A technical detail is deferred to the
Appendix.

\section{RPA Equations}
\label{rpa}

For the sake of completeness, we briefly recall the derivation of the
RPA equations for particle-hole pairs of Fermions. The derivation uses
the equations--of--motion approach. We follow Rowe's book~\cite{Row}.
It is essentially assumed that pairs of Fermions can approximately be
considered as Bosons.

We thus consider $N$ Bosonic single--particle states with creation
operators $B^{\dag}_k$ and annihilation operators $B^{}_k$ where $k =
1, \ldots, N$. These fulfill the commutation relations
\be
[ B^{}_k, B^{\dag}_l] = \delta_{k l} \ .
\label{1}
\ee
We write the Hamiltonian in a form which does not conserve particle
number and which is fully analogous to the Hartree--Fock--BCS
Hamiltonian for Fermions,
\be
H = \sum_{k l} A^0_{k l} B^{\dag}_k B^{}_l + \frac{1}{2} \sum_{k l}
C_{k l} B^{\dag}_k B^{\dag}_l + \frac{1}{2} \sum_{k l} C^*_{k l}
B^{}_k B^{}_l \ .
\label{2}
\ee
The matrices $A^0$ and $C$ both have dimension $N$. For the matrices
$A^0$ and $C$ we consider two options. The matrix $A^0$ may be
Hermitean or real symmetric. Since the $B$s commute, the matrix $C$ is
symmetric and either complex or real. Thus,
\ba
A^0 &=& (A^0)^{\dag} \ , \ C^{} = C^{\rm T} \ {\rm (``unitary \ 
case''), \ or} \nonumber \\
A^0 &=& (A^0)^* = (A^0)^T \ , \ C^{} = C^{\rm T} = C^* \ {\rm
(``orthogonal \ case'')} \ .
\label{3}
\ea
We observe that for Fermions, $C$ in Eq.~(\ref{2}) would be
antisymmetric. Let $| RPA \rangle$ denote the ground state of the
Hamiltonian $H$ in Eq.~(\ref{2}) with energy $E_0$,
\be
H |RPA\rangle = E_0 |RPA\rangle \ .
\label{6}
\ee
We postulate that the excited states of the system are created by
applying the operators
\be
Q^{\dag}_\nu = \sum_k \bigg( X^\nu_k B^{\dag}_k - Y^\nu_k B^{}_k
\bigg)
\label{4}
\ee
to the ground state $|RPA\rangle$,
\be
H Q^{\dag}_\nu |RPA\rangle = E_\nu |RPA\rangle \ .
\label{5}
\ee
Multiplying Eq.~(\ref{6}) with $Q^{\dag}_\nu$ and subtracting the
result from Eq.~(\ref{5}) we get
\be
[ H, Q^{\dag}_\nu ] |RPA\rangle = (E_\nu - E_0) |RPA\rangle \ .
\label{7}
\ee
The ground state is assumed to be annihilated by the operators
$Q^{}_\nu$ and by the operators
\be
\delta Q^{}_\rho = \sum_k \bigg( \delta X^{\rho *}_k B^{}_k - \delta
Y^{\rho *}_k B^{\dag}_k \bigg)
\label{8}
\ee
obtained by arbitrary variations of the $Q^{}_\nu$,
\be
\delta Q^{}_\rho |RPA\rangle = 0 \ {\rm for \ all} \ \rho \ .
\label{9}
\ee
Multiplication of Eq.~(\ref{7}) from the left by $\delta Q^{}_\rho$ 
and use of Eq.~(\ref{9}) yields
\be
[ \delta Q^{}_\rho , [ H , Q^{\dag}_\nu ] ] |RPA\rangle = (E_\nu -
E_0) [ \delta Q^{}_\rho, Q^{\dag}_\nu ] |RPA\rangle \ .
\label{10}
\ee
Working out the commutators, using the fact that the coefficients
$\delta X^{\rho *}_k$ and $\delta Y^{\rho *}_k$ are completely
arbitrary, and using the first of Eqs.~(\ref{3}), we obtain
\be
{\cal H}^0 {\vec X}^{\nu {\rm T}} = (E_\nu - E_0) {\vec X}^{\nu {\rm
T}} \ .
\label{14}
\ee
We have defined
\be
{\cal H}^0 = \left( \matrix{ A^0 & C \cr -C^* & -(A^0)^* \cr} \right) 
\label{12}
\ee
and
\be
{\vec X}^\nu = ( X^\nu_k , Y^\nu_k ) \ . 
\label{13}
\ee
Eqs.~(\ref{14}) to (\ref{13}) are the RPA equations for Fermionic
particle--hole pairs. Except for one important difference these
equations look very similar to the Hartree--Fock--BCS equations for
Fermions: The matrix $C$ is symmetric while in the BCS case, the
pairing potential is antisymmetric. As a consequence the BCS
Hamiltonian is Hermitean while the matrix ${\cal H}^0$ in
Eq.~(\ref{12}) is not. This is why the 2N $\times$ 2N matrix ${\cal
H}^0$ does not belong to one of the symmetry classes studied in
Ref.~\cite{Hei05}.

\section{Spectral Properties}
\label{ex}

Since the RPA matrix ${\cal H}^0$ is not Hermitean, the eigenvalues
need not be real. We investigate some of the consequences.

The matrix ${\cal H}^0$ in Eq.~(\ref{12}) has the following two
symmetry properties. With ${\bf 1}_N$ the unit matrix in $N$
dimensions and
\ba
{\cal M}_1 &=& \left( \matrix{ 0 & {\bf 1}_N \cr {\bf 1}_N & 0 \cr}
\right) \ , \nonumber \\
{\cal M}_2 &=& \left( \matrix{ {\bf 1}_N & 0 \cr 0 & -{\bf 1}_N \cr}
\right) \ ,
\label{13b}
\ea
we have
\ba
{\cal M}_1 ({\cal H}^0)^* {\cal M}_1 &=& - {\cal H}^0 \ , \nonumber \\
{\cal M}_2 ({\cal H}^{0})^{\dag} {\cal M}_2 &=& {\cal H}^{0} \ .
\label{11}
\ea
Let $E_\mu$ be an eigenvalue and $\Psi_\mu$ be an eigenfunction of the
RPA matrix,
\be
{\cal H}^0 \Psi_\mu = E_\mu \Psi_\mu \ .
\label{13a}
\ee
From the first of Eqs.~(\ref{11}) and $({\cal M}_1)^2 = {\bf 1}_{2N}$
it follows that ($- E^*_\mu$) is also an eigenvalue with eigenfunction
${\cal M}_1 \Psi^*_\mu$. From the second of Eqs.~(\ref{11}) and
$({\cal M}_2)^2 = {\bf 1}_{2N}$ it follows that $E^*_\mu$ is also an
eigenvalue with left eigenfunction $\Psi^{\dag}_\mu {\cal M}_2$.  We
conclude that with $E_\mu$ also $E^*_\mu$, $- E_\mu$ and $- E^*_\mu$
are eigenvalues of the RPA matrix ${\cal H}^{(0)}$. In other words,
real and purely imaginary eigenvalues come in pairs with opposite
signs, while fully complex eigenvalues with $\Re E_\mu \neq 0$ and
$\Im E_\mu \neq 0$ come in quartets that are symmetric with respect to
both the real and the imaginay energy axis.

For $C = 0$ the matrix ${\cal H}^0$ is Hermitean or real symmetric.
Then the eigenvalues of ${\cal H}^0$ are real and appear in pairs with
opposite signs. While level repulsion is the hallmark of a Hermitean
Hamiltonian, we show that the non--Hermitean part $C$ of ${\cal H}^0$
leads to repulsion (attraction) of levels with the same (opposite)
signs, respectively. With increasing strength of $C$ level attraction
leads to coalescence of pairs of eigenvalues with opposite signs at
energy $E = 0$. Coalescence is a signal for the instability of the RPA
approach. With a further increase of the strength of $C$ the two
coalesced eigenvalues leave the point $E = 0$ in opposite directions
along the imaginary $E$--axis. On the imaginary axis, a pair of
eigenvalues may coalesce again and then move away from the imaginary
axis in opposite directions along a line that is parallel to the real
$E$--axis. Because of the symmetry of the RPA matrix, such behavior
must occur in parallel on the positive and on the negative imaginary
axis.

Before giving the general argument, we demonstrate these statements
using two simple examples. To be specific we consider the unitary
case. First, we take $N = 1$ and label the matrix elements in an
obvious way by $a$ and $c$, respectively.  Here $a$ is real and $c$ is
complex. The eigenvalue equation is quadratic with the two solutions
$\Omega = \pm \sqrt{a^2 - |c|^2}$.  Increasing the value of $|c|$, we
pull the two eigenvalues toward each other. The two eigenvalues
coalesce at $E = 0$ for $|a| = |c|$ and become purely imaginary for
$|a| < |c|$. Second, we take $N = 2$. We assume that $A^0$ has been
diagonalized with eigenvalues $E_1$ and $E_2$ where, in the spirit of
the RPA, we assume $0 < E_1 < E_2$. The complex matrix elements of $C$
are labelled $c_1, c_2, c_{12}$ in an obvious way. The secular
equation is
\ba
&&(\Omega^2 - E^2_1)(\Omega^2 - E^2_2) + |c_1|^2 (\Omega^2 - E^2_2) +
|c_2|^2 (\Omega^2 - E^2_1) \nonumber \\
&& \qquad + 2 |c_{12}|^2 (\Omega^2 - E_1 E_2) + |\det(C)|^2 = 0
\label{14a}
\ea
which we write in the form
\ba
&& \Omega^4 - \Omega^2 (T_A - T_C) + D_C + E^2_1 E^2_2 \nonumber \\
&& \qquad \qquad = E^2_1 |c_2|^2 + E^2_2 |c_1|^2 + 2 E_1 E_2
|c_{1 2}|^2 \nonumber \\
&& \qquad \qquad = T_{AC} \ .
\label{14b}
\ea
Here $T_A = E^2_1 + E^2_2$, $T_C = {\rm Trace} [ C C^* ]$, $D_C =
|\det C|^2$, and the last part of Eq.~(\ref{14b}) defines $T_{AC}$. 
This quadratic equation for $\Omega^2$ has the solutions
\be
\Omega^2 = \frac{1}{2} (T_A - T_C) \pm \frac{1}{2} \sqrt{(T_A -
T_C)^2 + 4 T_{AC} - 4 E^2_1 E^2_2 - 4 D_C} \ .
\label{14c}
\ee
The argument of the square root vanishes for
\be
D_C = T_{AC} + \frac{1}{4} (T_A - T_C)^2 - E^2_1 E^2_2 
\label{14d}
\ee
while two eigenvalues coincide at $E = 0$ if
\be
D_C = T_{AC} \ .
\label{14e}
\ee
By definition we have $T_A > 0$, $T_C \geq 0$, $T_{AC} \geq 0$ and
$D_C \geq 0$. Inspection shows that also $(T_A - T_C)^2 - 4 E^2_1
E^2_2 \geq 0$. As we turn on the coupling matrix $C$, $T_C, T_{AC}$,
and $D_C$ grow monotonically from zero, and the condition~(\ref{14e})
is met prior to condition~(\ref{14d}). This shows that the smaller
of the two positive eigenvalues coalesces with its negative
counterpart at zero (and becomes imaginary) without the two positive
(or the two negative) eigenvalues ever coalescing.

We show that coalescence of two eigenvalues of opposite signs at $E =
0$ (without coalescence of pairs of positive or pairs of negative
eigenvalues) is a general property of the RPA equations. We denote the
orthogonal projectors onto the two $N$--dimensional subspaces
appearing explicitly on the right--hand side of Eq.~(\ref{12}) by
$Q_1$ and $Q_2$, respectively. With $E$ the energy, we eliminate the
subspace with projector $Q_2$ at the expense of introducing into the
first subspace the effective Hamiltonian $H = A^0 - C (E +
A^{0*})^{-1} C^*$. The term added to $A^0$ is a Hermitean (or a real
symmetric) operator and, thus, causes level repulsion between the
eigenvalues of $A^0$ which, therefore, cannot coalesce as the matrix
elements of $C$ increase in magnitude. (The non--Hermiticity of the
matrix ${\cal H}^0$ leads to the negative sign of the term $- C (E +
A^{0*})^{-1} C^*$, while a Hermitean ${\cal H}^0$ would have given a
positive sign. But level repulsion is independent of the signs of the
matrix elements connecting the states). Going to the eigenvalue
representation of the matrix $A$ and assuming that all eigenvalues are
positive, we find that the trace of $- C (E + A^{0*})^{-1} C^*$ is
negative for all $E \geq 0$ and largest for $E = 0$. This shows that
with increasing strength of $C$ the center of the spectrum of the
positive eigenvalues is shifted towards smaller values, the shift
being strongest for the levels near $E = 0$. We conclude that the
interaction described by $C$ causes level attraction between the
eigenvalues of $A^0$ and those of $- A^{0*}$ while it causes level
repulsion amongst the eigenvalues of $A^0$ and amongst those of
$A^{0*}$.

To investigate the coalescence of eigenvalues at $E = 0$ we follow
Kato's book~\cite{Kat66}. We denote by ${\cal H}^0(z)$ the matrix
obtained from ${\cal H}^0$ by the replacement $C \to zC$. We assume
that the eigenvalues of $A$ are all positive. We consider the
characteristic polynomial $P(\Omega, z)$ $= \det [ {\cal H}^0(z) -
\Omega {\bf 1}_{2 N} ]$ for complex values of $z$. The following
statements follow directly from Chapter II.1 of Ref.~\cite{Kat66}: The
roots $\Omega(z)$ of the equation $P(\Omega, z) = 0$ are branches of
analytic functions of $z$ with only algebraic singularities. At the
point $z = z_c$ where two or more eigenvalues coalesce two eigenvalues
assume the common value $\Omega_c$ if the two equations
\be
P(\Omega_c, z_c) = 0
\label{14f}
\ee
and
\be
\frac{\partial}{\partial \Omega} P(\Omega, z_c)|_{\Omega = \Omega_c}
\label{14g}
\ee
are both fulfilled. For three or more eigenvalues to coalesce, further
equations involving higher derivatives of $P$ must also be fulfilled.
Such equations impose constraints on ${\cal H}^0$. Therefore, the
generic case is the coalescence of two eigenvalues. (This is a
generalization~\cite{Sch73} of the Wigner--von Neumann non--crossing
rule for eigenvalues). We confine ourselves to that case. Two
eigenvalues coalescing at $z_c$ generically possess a square--root
branch point at $z_c$ (the case where the two coalescing eigenvalues
are holomorphic at $z_c$ also leads to constraints on ${\cal H}^0$, is
non--generic and, thus, likewise excluded~\cite{Sch73}). The same
argument applies to the coalescence of eigenvalues on the positive and
on the negative imaginary $E$--axis.

The behavior of two coalescing eigenvalues was already studied in the
simplest case $N = 1$ introduced above. Both eigenvalues are real and
have opposite signs for $z < z_c = |a|/|c|$ and are purely imaginary
and have opposite signs for $z > z_c$. With increasing real $z < z_c$
the two eigenvalues move along the real axis towards each other until
they coalesce at $z = z_c$. As $z$ increases further, the two
eigenvalues leave the point zero and move along the imaginary axis in
opposite directions.

That behavior is not confined to the case $N = 1$ but represents the
generic situation. We consider a point of coalescence $z_c$ with $z_c$
real where the two coalescing eigenvalues (denoted by $\Omega_1$ and
$\Omega_2$) both have the value zero. We assume that for $z$ real and
$z < z_c$, $\Omega_1$ ($\Omega_2$) is positive (negative,
respectively). For $z$ in the vicinity of $z_c$ we therefore must have
\be
\Omega_{1, 2} = \pm \beta \sqrt{z_c - z}
\label{14h}
\ee
where $\beta$ is some positive constant. Eq.~(\ref{14h}) shows that
while $\Omega_1$ ($\Omega_2$) is positive (negative) for $z$ real and
$z < z_c$, both $\Omega_1$ and $\Omega_2$ are purely imaginary and
carry opposite signs for $z$ real and $z > z_c$.

The case of two eigenvalues coalescing on the imaginary axis is
similar. With $z_c$ the critical strength for coalescence at $\Omega =
\Omega_c$ with $\Omega_c$ purely imaginary, the two eigenvalues are
purely imaginary for $z < z_c$. Therefore, Eq.~(\ref{14h}) takes the
form $\Omega_{1, 2} = \Omega_c \pm \beta \sqrt{z_c - z}$ where $\beta$
is purely imaginary. It follows that for $z > z_c$ the two eigenvalues
leave the imaginary axis in opposite directions along a straight line
that runs parallel to the real $E$--axis.

\section{Random--Matrix Approach}
\label{rma} 

In order to study the generic properties of the RPA equations, we
consider a Gaussian random--matrix model. In Fermionic systems, the
RPA equations~(\ref{14}) to (\ref{13}) account for the particle--hole
interaction. As that interaction tends to zero, the matrix $C$
vanishes while the matrix $A^0$ tends to a diagonal matrix with
elements given by the single--particle energies of the particle--hole
pairs. We assume that all these energies have the same value which we
denote by $r$. In the nuclear case we have $r = \hbar \omega$ where
$\omega$ is the frequency of the harmonic--oscillator potential. We
write $A^0$ in the form
\be
A^0 = r {\bf 1}_N + A \ .
\label{15}
\ee
The Hermitean (or real symmetric) N--dimensional matrix $A$ and the
complex (or real) symmetric matrix $C$ represent the particle--hole
interaction. We assume $A$ to be a member of the Gaussian unitary
ensemble (GUE) or of the Gaussian orthogonal ensemble (GOE), as the
case may be. The matrix elements are complex or real
Gaussian--distributed random variables with mean values zero and
second moments given by
\ba
\langle A_{\mu \nu} A_{\rho \sigma} \rangle &=& \frac{\lambda^2}{N}
\delta_{\mu \sigma} \delta_{\nu \rho} \ {\rm (unitary \ case), \ or}
\nonumber \\
\langle A_{\mu \nu} A_{\rho \sigma} \rangle &=& \frac{\lambda^2}{N}
(\delta_{\mu \sigma} \delta_{\nu \rho} + \delta_{\mu \rho}
\delta_{\nu \sigma}) \ {\rm (orthogonal \ case)} \ .
\label{16}
\ea
The indices run from $1$ to $N$. The angular brackets denote the
ensemble average. For $N \gg 1$, the average GUE and GOE level
densities have the shape of a semicircle with radius $2 \lambda$. The
GUE (GOE) is invariant under unitary (orthogonal) transformations $A
\to U A U^{\dag}$ ($A \to O A O^T$) where $U U^{\dag} = {\bf 1}_N$
($O O^T = 1_N$, respectively). The elements of $C$ are assumed to be
Gaussian--distributed random variables with zero mean values and
second moments
\ba
\langle C_{\mu \nu} C^*_{\rho \sigma} \rangle &=& \frac{\gamma^2}{N}
\bigg( \delta_{\mu \sigma} \delta_{\nu \rho} + \delta_{\mu \rho}
\delta_{\nu \sigma} \bigg) \ , ~
\langle C_{\mu \nu} C_{\rho \sigma} \rangle = 0 \ {\rm (unitary \
case)} \ , \nonumber \\
\langle C_{\mu \nu} C_{\rho \sigma} \rangle &=& \frac{\gamma^2}{N}
\bigg( \delta_{\mu \sigma} \delta_{\nu \rho} + \delta_{\mu \rho}
\delta_{\nu \sigma} \bigg) \ {\rm (orthogonal \ case)} \ .
\label{17}
\ea
The ensemble of matrices $C$ is invariant under the transformations $C
\to U C U^T$ where $U U^{\dag} = {\bf 1}_N$ and where $T$ denotes the
transpose. The elements of $A$ and of $C$ are assumed to be
uncorrelated. The spectrum of ${\cal H}$ is expected to be real as
long as the matrix elements of $C$ are not too large. This is why we
parametrized the Gaussian distribution of $C$ in terms of an extra
parameter $\gamma$. We simplify the presentation by considering only
the unitary case in the sequel. The orthogonal case is obtained with a
slight change of notation; our conclusions apply to both cases.

It is useful to display the statistical properties of $A$ and of $C$
in some detail. Writing the elements of the matrix $A$ in the form
\be
A_{\mu \nu} = \Re A_{\mu \nu} + i \Im A_{\mu \nu}
\label{16a}
\ee
we note that $\Re A_{\mu \nu}$ is real symmetric and $\Im A_{\mu \nu}$
is real antisymmetric. The elements of $\Re A_{\mu \nu}$ and of $\Im
A_{\mu \nu}$ are uncorrelated Gaussian--distributed random variables
with zero mean value and second moments given by
\ba
\langle ( \Re A_{\mu \nu} )^2 \rangle &=& (1 + \delta_{\mu \nu})
\frac{\lambda^2}{2 N} \ , \nonumber \\
\langle ( \Im A_{\mu \nu} )^2 \rangle &=& (1 - \delta_{\mu \nu})
\frac{\lambda^2}{2 N} \ .
\label{16b}
\ea
Similarly, writing the elements of the matrix $C$ as
\be
C_{\mu \nu} = \Re C_{\mu \nu} + i \Im C_{\mu \nu}
\label{16c}
\ee
we note that both $\Re C_{\mu \nu}$ and $\Im C_{\mu \nu}$ are real
symmetric. The elements of $\Re C_{\mu \nu}$ and of $\Im C_{\mu \nu}$
are uncorrelated Gaussian--distributed random variables with zero
mean value and second moments given by
\ba
\langle ( \Re C_{\mu \nu} )^2 \rangle &=& (1 + \delta_{\mu \nu})
\frac{\gamma^2}{2 N} \ , \nonumber \\
\langle ( \Im C_{\mu \nu} )^2 \rangle &=& (1 + \delta_{\mu \nu})
\frac{\gamma^2}{2 N} \ .
\label{16d}
\ea

With these definitions, the RPA matrix ${\cal H}^0$ takes the form
\ba
{\cal H}^0 &=& \left( \matrix{ r{\bf 1}_N & 0 \cr 0 & - r{\bf
1}_N \cr} \right) + \left( \matrix{ A & C \cr -C^* & -A^* \cr} \right)
\nonumber \\
&=& \left( \matrix{ r {\bf 1}_N & 0 \cr 0 & - r {\bf 1}_N \cr} \right)
+ {\cal H} \ . 
\label{18}
\ea
We refer to the joint invariance properties of $A$ and $C$ as to
``generalized unitary invariance'' (``generalized orthogonal
invariance'', respectively). The generalized unitary invariance says
that the ensemble of matrices ${\cal H}$ is invariant under the
transformation
\be
{\cal H} \to \left( \matrix{ U & 0 \cr 0 & U^* \cr} \right) {\cal H}
\left( \matrix{ U^{\dag} & 0 \cr 0 & U^T \cr} \right) \ .
\label{18a}
\ee
It was mentioned already that since ${\cal H}$ is not Hermitean, the
random--matrix ensemble constructed above does not belong to one of
the classes listed in Ref.~\cite{Alt}. Non--Hermitean or Non--Cartan
random--matrix models have recently received some attention, see
Refs.~\cite{Mag,Ber}. It seems that our model appears in Table 7 of
Ref.~\cite{Mag} as class 21b (lower sign). We are not aware, however,
of any detailed analysis of the model.

\begin{figure}[!h]
\centering
\epsfig{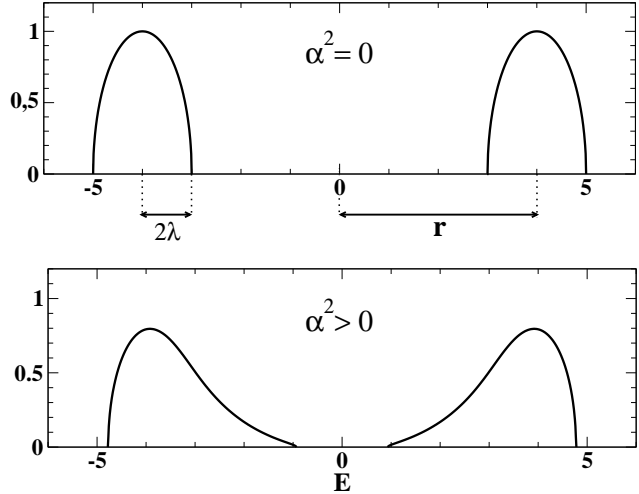}
\caption{Expected deformations and shifts of the two branches of the
average spectrum with increasing $\alpha^2$ (schematic).}
\label{fig0}
\end{figure}

We study our random--matrix models in the limit $N \to \infty$. In
that limit, the parameter $N$ disappears, and each of our
random--matrix models is characterized by two parameters. 
We introduce the first by noting that for $\gamma = 0$
the average spectrum consists of two semicircles. Each of these has
radius $2 \lambda$. We measure all energies in units of that radius.
The first parameter $r$ is then the distance of the center of each
semicircle from the origin at $E = 0$. The second parameter measures
the strength $\gamma$ of the matrix $C$ in units of the strength
$\lambda$ of the matrix $A$, see Eqs.~(\ref{16}) and (\ref{17}). It is
given by
\be
\alpha^2 = \frac{\gamma^2}{\lambda^2} \ .
\label{18c}
\ee
We will use these parameters when we treat the Pastur equation. Using
the results of Section~\ref{ex} we anticipate that with increasing
$\alpha^2$, the two branches of the average spectrum are deformed and
move toward each other. This is indicated schematically in
Fig.~\ref{fig0}.

\begin{figure}[!h]
\centering
\epsfig{file=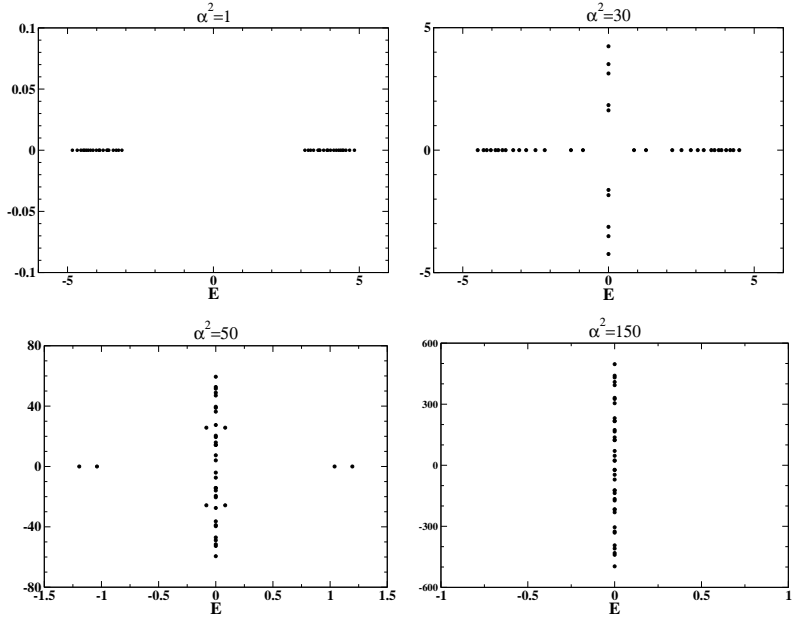,width=0.75\textwidth}
\caption{Evolution of the location in the complex energy plane of the
eigenvalues for $r = 4$ and $2N = 40$ when the coupling is increased,
i.e. for $\alpha^2 = 1, 30, 50, 150$.}
\label{fig0a}
\end{figure}

The average spectrum of our random--matrix model~(\ref{18}) takes a
simple form in two limiting cases. Aside from the simplification
encountered for $C = 0$ and discussed above, another simplification
arises in the opposite extreme case when the spectrum is dominated by
the matrix $C$. To see this we put in Eq.~(\ref{18}) $r = 0$, $A = 0$
and $C = i D$ and write the eigenvalues as $i \Omega$. The resulting
eigenvalue equation is
\be
\left( \matrix{ 0 & D \cr
                D^* & 0 \cr} \right) \Psi = \Omega \Psi \ .
\label{18d}
\ee
That equation possesses chiral symmetry. It is straightforward to
derive the Pastur equation and to see that the average spectrum has
the shape of a semicircle centered at zero with radius $2 \lambda$. We
conclude that as $\alpha^2$ in the RPA matrix in Eq.~(\ref{18})
increases from zero to very large values, the eigenvalues move from
the two semicircles on the real axis onto a single semicircle on the
imaginary axis. As they do so, they may intermittently leave both the
real and the imaginary $E$--axis and form groups of quartets arranged
symmetrically with respect to both these axes. This is exemplified in
Fig.~\ref{fig0a}. For $r = 4$ and several values of $\alpha^2$, we show
the distribution of eigenvalues in the complex energy plane. These
were obtained by diagonalization of randomly drawn matrices of
dimension $2 N = 40$. We see that before all eigenvalues reach the
imaginary axis at $\alpha^2 = 150$, a quartet is formed at $\alpha^2 =
50$. We have found numerous such quartets in other random
realizations. Our simulations also suggest that in the unitary case,
the semicircle formed for very large values of $\alpha^2$ on the
imaginary axis has a gap at $E = 0$. In certain chiral ensembles, the
average level density is known~\cite{Amb,Ake} to vanish at $E = 0$,
and our result probably relates to this fact. We have not followed
this up any further.

\section{Pastur Equations}
\label{pas} 

We ask: How do the matrix elements of $C$ affect the average RPA
spectrum? Which is the smallest value of $\gamma$ where the average
RPA spectrum becomes unstable? We answer these questions by
calculating the average retarded Green's function of the system. It is
given by
\be
\langle G(E) \rangle = \bigg\langle \bigg( E^+ {\bf 1}_{2N} - \left(
\matrix{ r{\bf 1}_N & 0 \cr 0 & -r{\bf 1}_N \cr} \right) - {\cal H}
\bigg)^{-1} \bigg\rangle \ .
\label{19}
\ee
For sufficiently small values of $\gamma^2$, we expect the average
spectrum of ${\cal H}^0$ to be real and to consist of two branches,
one at positive and one at negative energies. The instability of the
RPA occurs when the two branches begin to touch.

To calculate the ensemble average, we expand $G(E)$ in powers of
${\cal H}$. This yields
\be
G(E) = G_0(E) + \sum_{n = 1}^{\infty} G_0(E) ( {\cal H} G_0(E) )^n
\ .
\label{20}
\ee
Here $G_0(E)$ is the Green's function of the system for ${\cal H} =
0$,
\be
G_0(E) = \bigg( E^+ {\bf 1}_{2N} - \left( \matrix{ r{\bf 1}_N & 0
\cr 0 & -r{\bf 1}_N \cr} \right) \bigg)^{-1} \ .
\label{20a}
\ee
For $N \gg 1$ the calculation of the ensemble average of $G$ is
standard: We average each term in the sum in Eq.~(\ref{20}) separately
using Wick contraction. For $N \gg 1$, we keep in each term only the
contributions with the maximum number of free summations over the $N$
level indices. The resulting series can be resummed and yields the
Pastur equation
\be
\langle G(E) \rangle = G_0(E) + G_0(E) \langle {\cal H} \langle G(E)
\rangle {\cal H} \rangle \langle G(E) \rangle \ .
\label{21}
\ee
Using the same notation as in Section~\ref{ex}, we introduce the
orthogonal projection operators $Q_i$, $i = 1,2$ onto the two
$N$--dimensional subspaces displayed explicitly on the right--hand
side of Eq.~(\ref{18a}), with $Q^2_1 = Q_1, Q_1 Q_2 = 0, Q_1 + Q_2 =
{\bf 1}_{2N}$ etc. The average Green's function is written as
\be
\langle G(E) \rangle = \sum_{i,j = 1}^2 Q_i \langle G(E) \rangle Q_j \ .
\label{21a}
\ee
The terms with $i \neq j$ on the right--hand side of Eq.~(\ref{21a})
vanish. Indeed, it takes an odd number of matrices $C$ and an even
number of matrices $C^*$ or an even number of matrices $C$ and an odd
number of matrices $C^*$ to connect subspaces 1 and 2. The elements of
$C$ being independent Gaussian random variables with mean values zero,
such terms vanish upon averaging. This leaves us with
\be
\langle G(E) \rangle = Q_1 \langle G(E) \rangle Q_1 + Q_2 \langle G(E)
\rangle Q_2 \ . 
\label{21b}
\ee
We project the Pastur equation~(\ref{21}) onto each of the two
subspaces, use Eq.~(\ref{21b}), observe that $G_0(E)$ commutes with
$Q_1$ and $Q_2$, and obtain
\ba
Q_1 \langle G(E) \rangle Q_1 &=& Q_1 G_0(E) Q_1 + (Q_1 G_0(E) Q_1)
\nonumber \\
&& \times \{ \langle (Q_1 {\cal H} Q_1) (Q_1 \langle G(E) \rangle Q_1)
(Q_1 {\cal H} Q_1) \rangle (Q_1 \langle G(E) \rangle Q_1) \nonumber \\
&&+ \langle (Q_1 {\cal H} Q_2) (Q_2 \langle G(E) \rangle Q_2) (Q_2
{\cal H} Q_1) \rangle (Q_1 \langle G(E) \rangle Q_1) \} \nonumber \\
&=& Q_1 G_0(E) Q_1 + (Q_1 G_0(E) Q_1) \nonumber \\
&& \times \{ \langle A (Q_1 \langle G(E) \rangle Q_1) A \rangle (Q_1 
\langle G(E) \rangle Q_1) \nonumber \\
&& - \langle C (Q_2 \langle G(E) \rangle Q_2) C^* \rangle (Q_1 \langle
G(E) \rangle Q_1) \}
\label{22}
\ea
and
\ba
Q_2 \langle G(E) \rangle Q_2 &=& Q_2 G_0(E) Q_2 + (Q_2 G_0(E) Q_2)
\nonumber \\
&& \times \{ \langle (Q_2 {\cal H} Q_2) (Q_2 \langle G(E) \rangle Q_2)
(Q_2 {\cal H} Q_2) \rangle (Q_2 \langle G(E) \rangle Q_2) \nonumber \\
&&+ \langle (Q_2 {\cal H} Q_1) (Q_1 \langle G(E) \rangle Q_1) (Q_1
{\cal H} Q_2) \rangle (Q_2 \langle G(E) \rangle Q_2) \} \nonumber \\
&=& Q_2 G_0(E) Q_2 + (Q_2 G_0(E) Q_2) \nonumber \\
&& \times \{ \langle A^* (Q_2 \langle G(E) \rangle Q_2) A^* \rangle
(Q_2 \langle G(E) \rangle Q_2) \nonumber \\
&& - \langle C^* (Q_1 \langle G(E) \rangle Q_1) C \rangle (Q_2 \langle
G(E) \rangle Q_2) \} \ .
\label{23}
\ea
We perform the averages over the matrices $A$ and $C$ using
Eqs.~(\ref{16}) and (\ref{17}), define for $i = 1,2$ the functions
\be
\sigma_i(E) = \frac{\lambda}{N} {\rm Trace} Q_i \langle G(E) \rangle
Q_i,
\label{24}
\ee
use that
\be
Q_1 G_0(E) Q_1 = \frac{1}{E - r} {\bf 1}_N \ , \ Q_2 G_0(E) Q_2 =
\frac{1}{E + r} {\bf 1}_N,
\label{25}
\ee
take the traces of Eqs.~(\ref{22}) and (\ref{23}) and obtain, using
$N \gg 1$,
\ba
\sigma_1 &=& \frac{\lambda}{E - r} + \frac{\lambda}{E - r} \lambda (
\sigma_1 - \frac{\gamma^2}{\lambda^2} \sigma_2 ) \lambda \sigma_1 \ , 
\nonumber \\
\sigma_2 &=& \frac{\lambda}{E + r} + \frac{\lambda}{E + r} \lambda (
\sigma_2 - \frac{\gamma^2}{\lambda^2} \sigma_1 ) \lambda \sigma_2 \ .
\label{26}
\ea
With $\alpha^2 = \gamma^2 / \lambda^2$ these equations can be rewritten
in the form
\ba
\sigma_1 &=& \frac{\lambda}{E - r - \lambda ( \sigma_1 - \alpha^2
\sigma_2 ) } \ , \nonumber \\
\sigma_2 &=& \frac{\lambda}{E + r - \lambda ( \sigma_2 - \alpha^2
\sigma_1 ) } \ .
\label{27}
\ea
These are two coupled quadratic equations in the unknown functions
$\sigma_1(E)$ and $\sigma_2(E)$. We refer to both Eq.~(\ref{21}) and
to Eqs.~(\ref{27}) as to the (generalized) Pastur equation(s). A
physical interpretation of the functions $\sigma_i(E)$ with $i = 1,2$
is given at the end of Section~\ref{symm}.

\section{Properties of $\langle G(E) \rangle$}
\label{symm} 

Before averaging, ${\rm Trace} \ G(E)$ can be expressed in terms of
the eigenvalues of ${\cal H}^0$. We confine ourselves to sufficiently
small values of $|\gamma|$ so that for $N \to \infty$, the average
spectrum of ${\cal H}^0$ consists of two separate branches, the
positive branch with eigenvalues $E_{\mu} > 0$, and the negative
branch with eigenvalues $- E_\mu < 0$. Here, $\mu = 1, \ldots, N$. We
have used the symmetries of the spectrum derived in Section~\ref{ex}.

We distinguish the advanced and the retarded Green's functions by
appropriate indices. With $E^\pm = E \pm i \delta$ where $E$ is real
and $\delta$ positive infinitesimal, we have
\ba
{\rm Trace} \ G(E)_{\rm ret, adv} &=& \sum_{i = 1}^2 \sum_{\mu =
1}^N \frac{1}{E^\pm + (-)^i E_{\mu}} \nonumber \\
&=& \int {\rm d} E' \ \frac{1}{E^\pm - E'} \sum_{i = 1}^2 \sum_{\mu =
1}^N \delta( E' + (-)^i E_{\mu} ) \ .
\label{35}
\ea
Eq.~(\ref{35}) is standard for Hermitean operators. But ${\cal H}$ is
not Hermitean. Therefore, we derive Eq.~(\ref{35}) in the Appendix.
Averaging over the ensemble converts the sum over delta functions into
the average level density $\rho(E)$,
\be
\rho(E) = \bigg\langle \sum_{i = 1}^2 \sum_{\mu = 1}^N \delta( E +
(-)^i E_{\mu} ) \bigg\rangle \ ,
\label{35a}
\ee
where $\rho(E)$ is normalized such that the integral over all energies
yields $2 N$. It is obvious that
\be
\rho(E) = \rho(- E) \ .
\label{40}
\ee
The average level density is a symmetric function of $E$. It differs
from zero within two energy intervals, one extending over positive
energies with end points, say, $E_a$ and $E_b$ where $E_b > E_a > 0$
and the other, extending over negative energies between $- E_b$ and $-
E_a$.

We use $\rho(E)$ to write $\langle G(E) \rangle$ in the form
\be
\langle {\rm Trace} \ G(E)_{\rm ret, adv} \rangle = \int {\rm d} E' \
\frac{\rho(E')} {E^\pm - E'} \ .
\label{36}
\ee
Eq.~(\ref{36}) shows that the analytic function $\langle {\rm Trace}
G(E) \rangle$ has two branch cuts. The branch cut at positive energy
extends from the branch point at $E = E_a$ to the branch point at $E =
E_b$. The branch cut at negative energies is the mirror image of the
first with respect to the origin. The discontinuity of $\langle {\rm
Trace} G(E) \rangle$ across the cut is given by
\be
\langle {\rm Trace} G(E)_{\rm adv} \rangle - \langle {\rm Trace}
G(E)_{\rm ret} \rangle = 2 i \pi \rho(E) \ . 
\label{37}
\ee
For Hermitean problems, the retarded and the advanced Green's
functions are related by complex conjugation. However, ${\cal H}$ is
not Hermitean, and $\langle G(E)_{\rm ret} \rangle$ and $ \langle
G(E)_{\rm adv} \rangle$ are not complex conjugates of each other.
Inspection of Eqs.~(\ref{19}) and (\ref{18}) shows that with ${\cal
M}$ defined in Eq.~(\ref{13b}) we have
\be
- {\cal M} \langle G^*_{\rm adv, ret}(- E) \rangle {\cal M} = \langle
G_{\rm adv, ret}(E) \rangle \ .
\label{19b}
\ee
As long as the two branches of the spectrum do not touch, the
functions $\sigma_1(E)$ and $\sigma_2(E)$ have a simple
physical interpretation. That interpretation uses Eq.~(\ref{24}) and
the spectral decomposition of $G(E)$ given in Eq.~(\ref{35}). We first
use the results in Section~\ref{ex} to conclude that if $\Psi_\mu$ is
a right eigenfunction of ${\cal H}^0$ with real eigenvalue $E_\mu >
0$, then $\Psi^{\dag}_\mu M_2$ is the left eigenfunction to the same
eigenvalue, while ${\cal M}_1 \Psi^*_\mu$ and $\Psi^T_\mu {\cal M}_1
{\cal M}_2$ are the right and left eigenfunctions, respectively, to
eigenvalue $- E_\mu$. We assume that the eigenvalues are not
degenerate. We expand $G(E)$ itself (rather than its trace) in the
form of Eq.~(\ref{35}) and project the result onto the subspaces with
index $1$ or $2$. We use ${\cal M}_2 Q_1 = Q_1$ and ${\cal M}_2 Q_2 =
- Q_2$. This yields
\ba
\Im Q_1 G(E)_{\rm ret} Q_1 &=& - \pi \sum_{\mu = 1}^N \langle
\Psi^{\dag}_\mu | Q_1 | \Psi_\mu \rangle \delta(E - E_\mu)
\nonumber \\
&& - \pi \sum_{\mu = 1}^N \langle \Psi^T_\mu | {\cal M}_1 Q_1 {\cal
M}_1 | \Psi^*_\mu \rangle \delta(E + E_\mu) \ , \nonumber \\
\Im Q_2 G(E)_{\rm ret} Q_1 &=& + \pi \sum_{\mu = 1}^N \langle
\Psi^{\dag}_\mu | Q_2 | \Psi_\mu \rangle \delta(E - E_\mu)
\nonumber \\
&& + \pi \sum_{\mu = 1}^N \langle \Psi^T_\mu | {\cal M}_1 Q_2 {\cal
M}_1 | \Psi^*_\mu \rangle \delta(E + E_\mu) \ .
\label{19c}
\ea
Eqs.~(\ref{19c}) show that $\sigma_1(E)$ ($\sigma_2(E)$) measures the
total average spectroscopic strength of all states in the subspace
with index $1$ (with index $2$, respectively). For $\alpha^2 = 0$,
$\sigma_1$ ($\sigma_2$) receives non--vanishing contributions from the
positive--energy states (the negative--energy states) only. As
$\alpha^2$ increases, that situation is expected to change. This is
confirmed by our numerical calculations.

The analytic properties of $\langle G(E) \rangle$ used in the present
Section may fail to hold when a finite fraction of eigenvalues (i.e.,
a fraction of order $N$) is neither purely real nor purely imaginary.
In that case use of the Pastur equations as in the present paper may
not give correct results because the Green's function has other
singularities besides branch points, see f.i. Ref.~\cite{Kol}.

\section{Properties of the Solutions of Eqs.~(\ref{27})}
\label{prop} 

We are going to use the solutions $\sigma_{1, 2}$ of Eqs.~(\ref{27})
to determine the average level density $\rho(E)$ of the system. By
elimination of $\sigma_2$, Eqs.~(\ref{27}) can be reduced to an
equation of fourth order for $\sigma_1$, see Eq.~(\ref{50}) below.
Thus, Eqs.~(\ref{27}) possess four pairs of solutions
$(\sigma^{(i)}_1, \sigma^{(i)}_2)$ $i = 1, \ldots, 4$. It is
neccessary to determine which of these is relevant to our problem and
how that solution relates to $\rho(E)$. We use perturbation theory for
small values of $\alpha^2$ and continuity to establish that connection.
In order to deal with non--overlapping spectra we take $2 \lambda < r$.

We first take $\alpha^2 = 0$. Then Eqs.~(\ref{27}) decay into two
uncoupled equations for the unperturbed spectroscopic strenghth
 $\Sigma_{1, 2}$. With
$\epsilon_i =(E + (-)^i r)/(2 \lambda)$ the equations read
\be
\Sigma_i = \frac{1}{2 \epsilon_i - \Sigma_i} \ , \ i = 1,2 \ .  
\label{E2}
\ee
The solutions are
\be
\Sigma_i = \epsilon_i \pm \sqrt{\epsilon^2_i - 1} \ , \ i = 1,2 \ .
\label{E3}
\ee
The solutions have square--root singularities (branch points) at
$\epsilon_i = \pm 1$. For $\Sigma_1$ ($\Sigma_2$), the two branch points
are at $E_a = r - 2 \lambda$ and at $E_b = r + 2 \lambda$ (at $- E_b$
and at $- E_a$, respectively). The branch cut extends along the real
$E$--axis from $E_a$ to $E_b$ (from $- E_b$ to $- E_a$, respectively)
and connects the two sheets of the Riemann surface. We define the
physical sheet as the one where the imaginary part of $\Sigma_i$ is
positive when we approach the branch cut from below the real axis.
Then the square root carries a negative sign for $E$ real and $E
\gg r$, and $\Sigma_i \approx \lambda / E$ for $E$ real and $|E| \gg
r$. On the second sheet we have $\Sigma_i \approx E / \lambda$ for $E$
real and $|E| \gg r$. For both solutions, the level density differs
from zero only if $E$ lies on the branch cut, see Eq.~(\ref{37}). In
both cases, the average level density is given by the value of $\Im
\Sigma_i$ on the physical sheet and has the form of Wigner's semicircle
law $(N / (\pi \lambda)) (1 - [(E - (\pm) r)/(2 \lambda)]^2)^{1/2}$.

We now consider the case of small $\alpha^2$ and solve
Eqs.~(\ref{27}) perturbatively up to first order in $\alpha^2$. This
yields
\ba
\sigma_1 &=& \Sigma_1 - \alpha^2 [\Sigma_1]^2 \Sigma_2 \ , \nonumber \\
\sigma_2 &=& \Sigma_2 - \alpha^2 [\Sigma_2]^2 \Sigma_1 \ .
\label{E5}
\ea
We note that the end points of the spectrum are not changed. We would
expect that with increasing $\alpha^2$, the branch points are also
shifted. Such a shift is, however, beyond the reach of a
straightforward perturbative approach, see the end of this Section. We
focus attention on $\sigma_1$ and consider the physical sheet of
$\Sigma_1$ where $\Im \Sigma_1 > 0$ for $E_a \leq E \leq E_b$. Because
of the symmetry of Eqs.~(\ref{E5}) our arguments carry over
straightforwardly to $\sigma_2$. The coupling between states in
subspaces 1 and 2 shifts some of the spectroscopic strength of
subspace 1 into subspace 2 and, thus, into the interval $- E_b \leq E
\leq - E_a$, and vice versa. Both this gain and the corresponding loss
of strength in the first interval are given by the last term in the
first of Eqs.~(\ref{E5}). The first of Eqs.~(\ref{19c}) shows that
(whatever choice we make for $\Im \sigma_1$) the contributions to $\Im
\sigma_1$ from the positive-- and from the negative--energy states
must have the same signs. We have chosen $\Im \Sigma_1 > 0$ for $E_a
\leq E \leq E_b$. Therefore, in the interval $- E_b \leq E \leq - E_a$,
the imaginary part of $\sigma_1$ given by $- \Im ([\Sigma_1]^2
\Sigma_2) = - [\Sigma_1]^2 \Im \Sigma_2$ must be likewise positive.
That means that $\Im \Sigma_2$ must have a negative sign, or that we
must choose $\Sigma_2$ to lie on the second sheet. We conclude that
$\Im \sigma_2$ must be negative: {\it The physically meaningful
solutions of Eqs.~(\ref{E5}) are the ones where $\Im \sigma_1$ and
$\Im \sigma_2$ carry opposite signs}. In the interval $E_a \leq E \leq
E_b$, the imaginary part of the last term in the first of
Eqs.~(\ref{E5}) has the form $- 2 \alpha^2 \Re \Sigma_1 \Im
\Sigma_1 \Sigma_2$. Here $\Sigma_2 \approx E / \lambda$ is positive on
the second sheet and grows monotonically with $E$, the positive
function $\Im \Sigma_1$ is given by the semicirle law and is symmetric
about $E = r$, and $\Re \Sigma_1$ is antisymmetric about that point
and negative for $E < r$. It follows that $- 2 \alpha^2 \Re \Sigma_1
\Im \Sigma_1 \Sigma_2$ is positive (negative) for $E < r$ ($E > r$,
respectively). Moreover, for equal distances of $E$ from $r$ the
negative part has larger magnitude than the positive one. Aside from
its loss to the negative energy interval, the level density in the
positive energy interval is shifted toward smaller energies. This was
expected, see Section~\ref{ex}. The total balance of all these changes
must be zero, and the energy integrals over $\Im \sigma_1$ and $\Im
\Sigma_1$ must be equal since $\int \rho(E) = 2 N$ by definition as
long as the two branches of the spectrum do not touch. We have not
checked that statement analytically for the terms in Eqs.~(\ref{E5})
but have tested our numerical results accordingly, see
Section~\ref{num}.

In summary we have shown that for small $\alpha^2$ we must choose
those solutions of Eqs.~(\ref{24}) for which the imaginary parts of
$\sigma_1$ and $\sigma_2$ have opposite signs. Using continuity we
conclude that this statement applies for all values of $\alpha^2$.
Eqs.~(\ref{24}) being real, the non--real solutions $(\sigma_1,
\sigma_2)$ come in complex conjugate pairs.  Without loss of
generality we can choose the solution for which $\Im \sigma_1$ is
positive. For $E \geq 0$ the total level density of the system is then
given by
\be
\rho(E) = + \frac{N}{\pi \lambda} \Im \big( \sigma_1(E) + \sigma_2(E)
\big) \ .
\label{E6}
\ee
We observe that Eq.~(\ref{E6}) and the symmetry of the system imply
that $\rho(E)$ vanishes at $E = 0$. This is true even when some of the
eigenvalues have moved off the real axis.

To investigate the motion of the branch points due to changing
$\alpha^2$, we use a modified version of perturbation theory. We
substitute in the first of Eqs.~(\ref{27}) for $\sigma_2$ the
asymptotic value $E / \lambda$ on the second sheet and obtain
\be
\sigma_1 = \frac{\lambda}{E(1 + \alpha^2) - r - \sigma_1} \ .
\label{32}
\ee
Solving this quadratic equation we find for the new branch points
$E'_a$ and $E'_b$ the values $E'_a = E_a / (1 + \alpha^2)$ and $E'_b =
E_b / (1 + \alpha^2)$. That shows that both branch points are shifted
toward smaller energies as expected.

\section{Numerical Results}
\label{num} 

By elimination of $\sigma_2$, we obtain from Eq.~(\ref{27}) an
equation for $\sigma_1$. It reads
\be
\label{50}
a\sigma_1^4+b\sigma_1^3+c\sigma_1^2+d\sigma_1+e=0
\ee
where 
\ba
a&=& \lambda^2 (1-\alpha^4) \ , \nonumber \\
b&=& \lambda(E-r)(\alpha^4-2)-\lambda \alpha^2(E+r) \ , \nonumber \\
c&=& (E-r)^2+\alpha^2(E^2-r^2)+2\lambda^2 \ , \nonumber \\
d&=& -2\lambda(E-r)-\lambda\alpha^2(E+r) \ , \nonumber \\
e&=& \lambda^2 \ .
\label{51}
\ea
Without loss of generality we put $2 \lambda = 1$. Equivalently, we go
over to dimensionless variables by measuring all energies in units of
$2 \lambda$. Then the semicircles have equal radii of value unity, $r$
is, for $\alpha^2 = 0$, the distance between the center of each of the
two semicircles and the origin, $\alpha^2$ the strength of the
coupling, and $E$ the dimensionless energy.

Except for the non--generic cases where the coefficients $a$ and $b$
are equal to zero, Eq.~(\ref{50}) has four solutions denoted by
$\sigma_1^{(i)}$ ($i=1,2,3,4$). We may either have $4$ complex
solutions (all four imaginary parts do not vanish identically), or $2$
complex and $2$ real solutions, or $4$ real solutions. Complex
solutions always come in complex conjugate pairs. To all four
solutions corresponds a unique value of $\sigma_2$ obtained from the
first (linear) equation of Eqs.~(\ref{27}). If $\sigma_1$ is real
(complex) then $\sigma_2$ is also real (complex). Thus we have the
same three possibilities for the pair ($\sigma_1^{(i)}$,
$\sigma_2^{(i)}$) : 4 real, 4 complex or 2 real and 2 complex
pairs. As explained in Section~\ref{prop} we look for the complex
solution with $\Im (\sigma_1) > 0$ and $\Im (\sigma_2) < 0$. It turns
out that that solution occurs for the cases when we have 2 real and 2
complex pairs of solutions, and when we have four complex solutions.

\begin{figure}[!hbt]
\centering
\epsfig{file=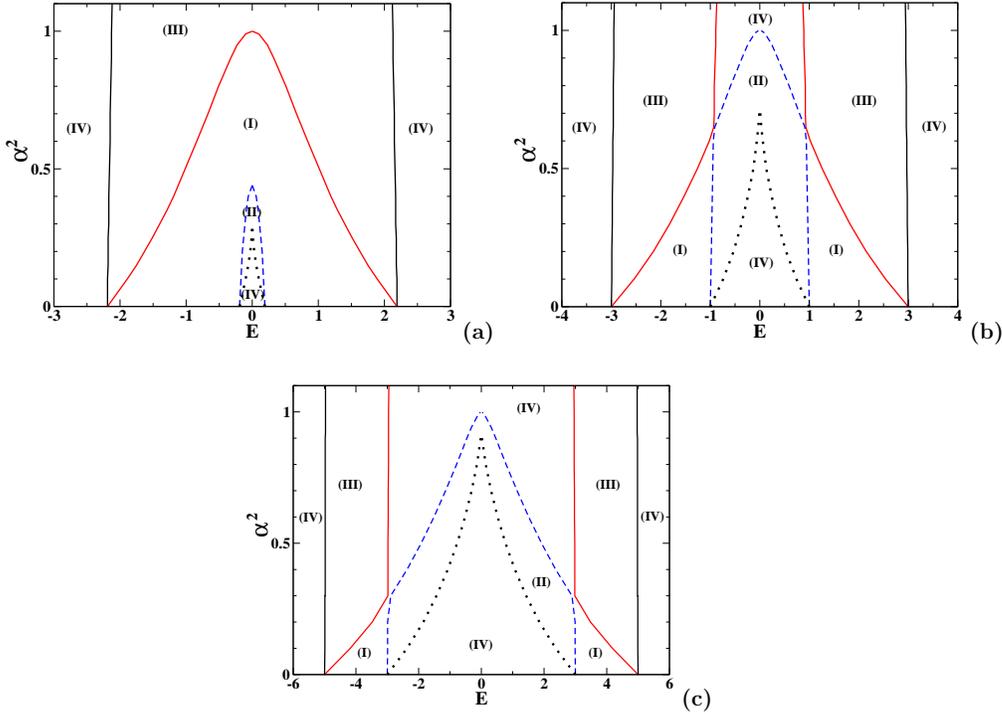,width=0.95\textwidth}
\caption{(Color online) The four domains (I) to (IV) in the 
($E$, $\alpha^2$) plane for three values of $r$. (a): r = 1.2, 
(b): r = 2.0, (c): r = 4.0 (See text for further explanation).}
\label{fig1}
\end{figure}

We summarize the properties of the solutions of Eqs.~(\ref{27}) found
numerically in Fig.~\ref{fig1}. For three values of $r$, we show in the
($E$, $\alpha^2$) plane four domains labelled (I) to (IV). These
symbols stand for
\begin{itemize}
\item (I) 4 complex pairs,
\item (II) 2 real and 2 complex pairs with equal signs for
$\Im(\sigma_1)$ and $\Im(\sigma_2)$,
\item (III) 2 real and 2 complex pairs with opposite signs for
$\Im(\sigma_1)$ and $\Im(\sigma_2)$,
\item (IV) 4 real pairs.
\end{itemize}
The physically interesting solutions lie in domains I and III. In all
three cases shown in Fig.~\ref{fig1}, the border lines between domains III
and IV are not vertical but slightly inclined: With increasing
$\alpha^2$, the spectra are pushed toward $E = 0$. Similarly, the
domains II and IV located at the center become squeezed and finally
disappear. For $\alpha^2 > 1$ we only have 4 real solutions (case IV)
or 2 real and 2 complex solutions whose imaginary parts have opposite
signs (case III).\\

\begin{figure}[!hbt]
\centering
\epsfig{file=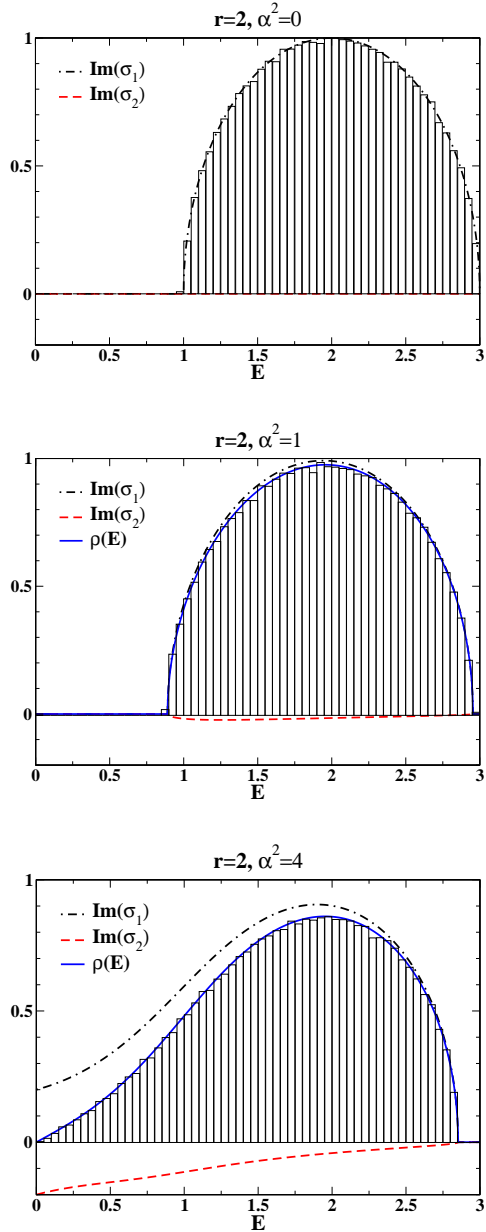,width=0.45\textwidth}
\caption{(Color online) The average level density $\rho (E)$ versus
$E$ for $E \geq 0$ as obtained from the Pastur equation. Black dashed--dotted 
line: $\Im\sigma_1$, red dashed line: $\Im \sigma_2$, blue solid line: sum of 
the two. For comparison, histograms corresponding to a numerical simulation with 
100 RPA matrices ${\cal H}$ of dimension $2N=100$, $r=2$ and different values
of $\alpha^2$.}
\label{fig1a}
\end{figure}

In Fig.~\ref{fig1a} we compare for $E \geq 0$ the average level
density obtained from the Pastur equation with the actual level
density obtained by numerical simulation. We have constructed 100
realizations of the RPA matrix in Eq.~(\ref{17}) with dimension $2 N =
100$ for the unitary case for $r = 2$ and three values of $\alpha^2$
by drawing the elements of the matrices $A$ and $C$ from Gaussian
distributions. We have determined the eigenvalues by diagonalization.
The results are shown as histograms. The imaginary parts of $\sigma_1$
and $\sigma_2$ are given as dot--dashes and dotted lines,
respectively, their sum as a solid line. The very close agreement
shows that the Pastur equation is capable of predicting average
spectra quite accurately even for matrices of fairly small dimension.

\begin{figure}[p]
\centering
\epsfig{file=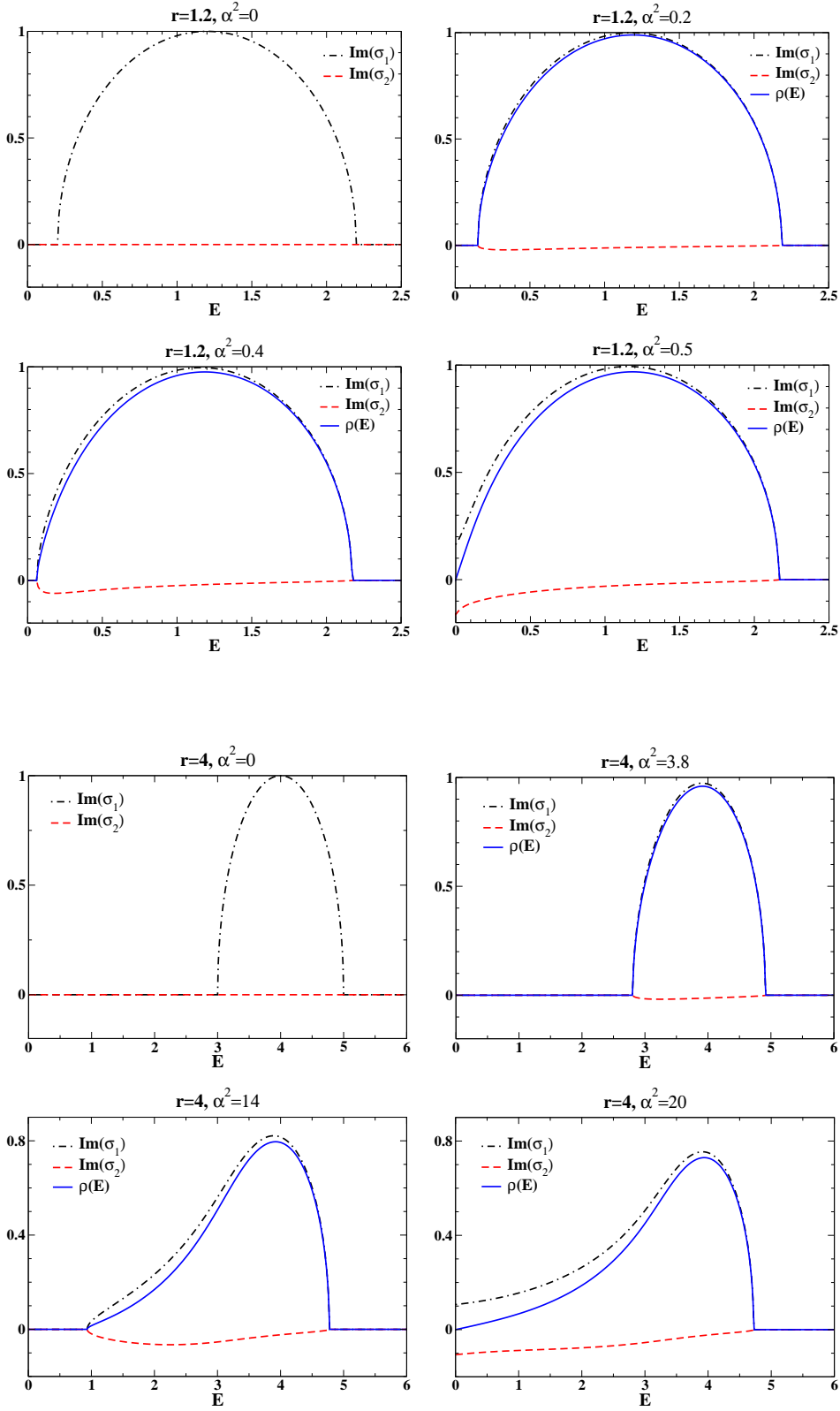,width=0.9\textwidth}
\caption{(Color online) The average level density $\rho(E)$ versus $E$ for $E
\geq 0$ as obtained from Eq.~(\ref{E6}). Upper (lower) panels:
$\rho(E)$ for $r = 1.2$ ($r = 4.0$) and for several values of the coupling 
parameter $\alpha^2$. Dahed--dotted, dashed and full lines as in Fig.~\ref{fig1a}.}
\label{fig2}
\end{figure}

Fig.~\ref{fig2} shows the level density $\rho(E)$ as obtained from
Eq.~(\ref{E6}) for $E \geq 0$. For $\alpha^2 = 0$ we obtain two
semicircles. These are separate for $r > 1$ but would overlap for
$r < 1$. With increasing $\alpha^2$, the semicircles become ever more
deformed and are pushed towards each other, as expected from the
perturbative result in Section~\ref{prop}. Average spectra that are
separate for $r > 1$ and $\alpha^2 = 0$ eventually touch. We observe
that the critical value of $\alpha^2$ where this happens (denoted by
$\alpha_{\rm crit}^2$) increases with increasing $r$, starting out from
$\alpha_{\rm crit}^2 = 0$ at $r = 1$.

\begin{figure}[!hbt]
\centering
\epsfig{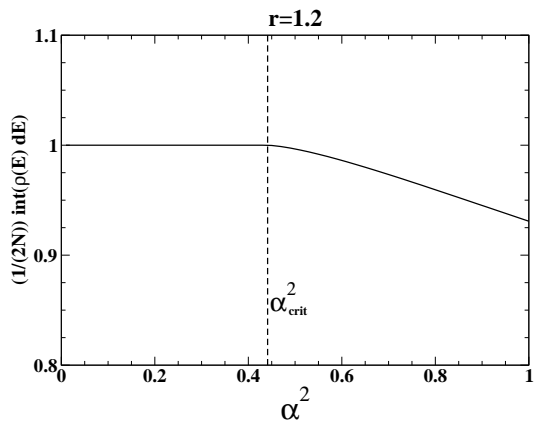}
\caption{The norm integral $(1 / (2 N)) \rho(E) \ \int {\rm d} E$ versus $\alpha^2$. 
See text for further explanation.}
\label{fig3}
\end{figure}

In Fig.~\ref{fig3} we show the value of the integral $I = (1 / (2 N))
\int {\rm d} E \ \rho(E)$ versus $\alpha^2$. For $\alpha^2 = 0$ the
integral has the value unity. For $0 < \alpha^2 \leq \alpha_{\rm crit}^2$
the value of $I$ should be very close to unity. In that regime, some
elements of the matrix $C$ for some rare realizations may have large
values so that some eigenvalues become non--real and do not contribute
to $I$. The resulting decrease of $I$ is expected to vanish as $N \to
\infty$, however, and is not visible in Fig.~\ref{fig3}. For $\alpha^2 >
\alpha_{\rm crit}^2$ the value of $I$ decreases monotonically with
$\alpha^2$ because an ever increasing macroscopic fraction of
eigenvalues leaves the real $E$--axis.

\section{Critical Strength}
\label{crit} 

The critical strength $\alpha_{\rm crit}^2$ is defined as the smallest
value of $\alpha^2$ for which for two solutions whose imaginary parts
carry different signs, $\Im \sigma_1 $ and $\Im \sigma_2 $ do not
vanish at $E = 0$. In order to find an analytical expression for
$\alpha_{\rm crit}^2$ as a function of $r$, we put in Eq.~(\ref{50})
$E = 0$ and define, using Eqs.~(\ref{51}), the new variable $y$ by
\ba
y = \sigma_1 +\frac{b}{4a} \ .
\label{53}
\ea
From Eq.~(\ref{50}) we obtain
\be
\label{54}
y^4 + uy^2 + vy+w = 0
\ee
where
\ba
\label{55}
u &=& -\frac{3b^2}{8a^2}+\frac{c}{a} \ , \nonumber \\
v &=& \frac{b^3}{8a^3}-\frac{bc}{2a^2}+\frac{d}{a} \ , \nonumber \\
w &=& -3\Bigl{(}\frac{b}{4a}\Bigr{)}^4+\frac{c}{a}\Bigl{(}\frac{b}{4a}
\Bigr{)}^2-\frac{bd}{4a^2}+\frac{e}{a} \ .
\ea
Eq.~(\ref{54}) is an equation of fourth order. The solutions are given
by a combination of the three roots of the third--order polynomial
$R(Z)$,
\be
\label{Rz}
R(Z)=Z^3+2uZ^2+(u^2-4w)Z-v^2 \ .
\ee
\noindent
The polynomial R(Z) has real coefficients. Therefore, the following
possibilities exist.
\begin{itemize}
\item (i) The three roots of $R(Z)$ are real and have equal signs;
the quartic equation~(\ref{54}) has 4 real solutions.
\item (ii) The three roots of $R(Z)$ are real, only two roots have the
same signs; the quartic equation~(\ref{54}) has 4 complex solutions.
\item (iii) Two roots of $R(Z)$ are complex and one is real; the
quartic equation~(\ref{54}) has 2 real and 2 complex solutions.
\end{itemize}
\noindent
The roots of $R(Z)$ are determined via the Cardan method. The
substitution
\ba
X = Z + \frac{2u}{3}
\ea
\noindent
eliminates the quadratic term in Eq.~(\ref{Rz}) and yields the
equation
\be
\label{56}
X^3+\tilde{u}X+\tilde{v} = 0
\ee
with
\ba
\tilde{u}&=&-\frac{1}{3}u^2-4w \ , \nonumber \\
\tilde{v}&=&\frac{2u}{27}(36w-u^2)-v^2 \ .
\label{57}
\ea
The nature of the roots is determined by the discriminant
\be
\label{discriminant}
\Delta=\tilde{v}^2+\frac{4}{27}\tilde{u}^3 \ .
\ee
\noindent
If $\Delta<0$ we have case (i), if $\Delta = 0$ we have case (ii), and
if $\Delta>0$ we have case (iii). Thus, $\alpha_{\rm crit}^2$ is given
by one of the two solutions of $\Delta = 0$. These are
\ba
(\alpha_{\rm crit}^{(1)})^2 &=& r^2-1 \ , \nonumber \\
(\alpha_{\rm crit}^{(2)})^2 &=& \frac{r^2-1}{r^2} \ .
\label{58}
\ea 
It turns out that 
$(\alpha_{\rm crit}^{(1)})^2$ ($(\alpha_{\rm crit}^{(2)})^2$) 
corresponds 
to the case where the imaginary parts of
$\sigma_1$ and $\sigma_2$ have different signs (the same sign,
respectively). We conclude that the physically interesting solution is
$(\alpha_{\rm crit}^{(1)})^2$ as given by the first of Eqs.~(\ref{58}).
For $r = 1$ (the two semicircles touch) it starts at unity and
increases with $r$.

\section{Summary and Conclusions}
\label{summ} 

In this paper, we have defined and investigated two random--matrix
models for the RPA equations. These models represent the generic forms
of the RPA equations in cases of unitary or orthogonal symmetry. The
RPA matrix is not Hermitean. Therefore, our random--matrix models do
not belong to one of the ten classes listed in Ref.~\cite{Hei05}. Both
random--matrix models depend on two parameters. The
parameter $\alpha^2$ gives the relative strength of the matrix $C$
connecting states at positive and states at negative energies in
relation to the strength of the matrix $A$, the matrix of interactions
amongst the positive--energy states. For $\alpha ^2= 0$, the average
spectrum has the shape of two semicircles, one at positive and one at
negative energies. The parameter $r$ gives the distance of the centers
of the two semicircles from the origin $E = 0$.

We have focussed attention on the forms of the average spectra. We
have not investigated the fluctuation properties of the positive-- and
the negative--energy branches of the spectra. This is because we are
confident that both branches display the usual GUE or GOE level
repulsion. Indeed, the total operator governing the positive--energy
branch of the spectrum is given by $A - \delta A$ where $\delta A = C
( E + r + A^* ) C^*$. The operator $A - \delta A$ depends on energy.
The addition of $\delta A$ to $A$ modifies the form of and shifts the
average spectrum. But when we consider some energy interval comprising
a finite number of levels in the limit $N \to \infty$, we may replace
the energy argument $E$ of $\delta A$ by the value of $E$ at the
center of that interval and take $\delta A$ as constant. The unitary
or orthogonal invariance of the random matrix $A - \delta A$ should
then suffice to guarantee local spectral fluctuation properties of the
GUE or GOE type.

While level repulsion dominates the positive-- and the
negative--energy branches of the spectrum, there is level attraction
between the two branches. That attraction is due to the matrix
$C$. The two semicircles describing the average spectrum for $C = 0$
become deformed and move toward each other as the strength $\alpha^2$ of
$C$ is increased. At some critical value $\alpha ^2= \alpha_{\rm crit}^2$,
the spectra touch at $E = 0$. That value marks the point where the RPA
equations become instable for most realizations of the random--matrix
models.

The mechanism that leads to instability is displayed by non--averaged
RPA spectra. The symmetry of the RPA matrix implies that eigenvalues
occur in pairs or in quartets: Real and purely imaginary eigenvalues
occur pairwise with opposite signs; for fully complex eigenvalues with
non--vanishing real and imaginary parts, the four eigenvalues lie
symmetrically with respect to both the real and the imaginary axis in
the complex $E$--plane. Level attraction leads to coalescence of pairs
of real eigenvalues with opposite signs at $E = 0$. With increasing
$\alpha^2$, the two eigenvalues leave the point $E = 0$ and move in
opposite directions along the imaginary axis. On that axis, pairs of
eigenvalues may coalesce again. Because of the symmetry of the RPA
equations, such coalescence must occur in parallel for the same value
of $\alpha^2$ on the positive and on the negative imaginary axis. A
further increase of $\alpha^2$ causes each pair to separate and to leave
the imaginary axis in opposite directions along a straight line that
runs parallel to the real $E$--axis. For very large values of
$\alpha^2$, all eigenvalues come to lie on the imaginary $E$--axis. The
average spectrum has the shape of a semicircle centered at $E = 0$.

We have studied these features both analytically and numerically. A
generalized Pastur equation was derived for large matrix dimension ($N
\to \infty$). It reduces to a fourth--order polynomial for the
spectral strength function $\sigma_1$ of the positive--energy
states. While for $\alpha^2 = 0$ $\Im \sigma_1$ is confined to positive
energies, $\Im \sigma_1$ spreads to negative energies as $\alpha^2$ is
increased. We have shown that the level density obtained from the
solution $\sigma_1$ agrees very well with the result of numerical
diagonalizations even of random RPA matrices of fairly low
dimension. From the fourth--order equation for $\sigma_1$ we derived
the critical value $\alpha_{\rm crit}^2$ of the strength $\alpha^2$ at
which the two branches of the spectrum touch each other.  That value
is given by the simple equation $\alpha^2_{\rm crit} = r^2 - 1$ where
$r$ is the distance of the center of each semicircle from the origin
$E = 0$ measured in units of the radius of the semicircle. That result
may be useful in applications: $\alpha^2$ may be estimated by the
ratio of two mean--square values measuring the strengths of the
interaction matrix elements connecting particle--hole states at
positive and negative energies (at positive energies only,
respectively). The parameter $r$ may be estimated as the
ratio of the mean value of the positive particle--hole energies and
the root--mean--square value of the interaction matrix elements
connecting states at positive energies.\\
\\
\textbf{Acknowledgements}: We thank G. Akemann and P. Brouwer for helpful
comments and for drawing our attention to Refs.~\cite{Mag,Ber,Kol} and 
M. R. Zirnbauer for clarifying to us the difference between 
the general Cartan classification and the Altland-Zirnbauer one.

\section*{Appendix}

We derive the eigenvalue decomposition~(\ref{35}) of the trace of the
Green's function. In Chapter 1, Eq.~(5.23) of Ref.~\cite{Kat66} it is
shown that the resolvent $R(\zeta) = (T - \zeta)^{-1}$ of an arbitrary
matrix $T$ in $N$ dimensions with eigenvalues $\lambda_h$, $h = 1, 2,
\ldots, s$ possesses the decomposition
\be
R(\zeta) = - \sum_{h = 1}^s \bigg[ (\zeta - \lambda_h)^{-1} P_h +
\sum_{n = 1}^{m_h - 1} (\zeta - \lambda_h)^{-n - 1} D_h^n \bigg] \ . 
\label{A1}
\ee
Here $\zeta$ is a complex variable and $\lambda_h \neq \lambda_k$ for
$h \neq k$. The matrices $P_h$ are projectors which obey (Eq.~(5.21)
of Ref.~\cite{Kat66})
\be
P_h P_k = \delta_{h k} P_h \ ; \ \sum_h P_h = 1 \ .
\label{A2}
\ee
The matrices $D_h$ obey (see Eqs.~(5.24) and (5.50) of
Ref.~\cite{Kat66})
\ba
P_h D_k &=& D_k P_h = \delta_{h k} D_k \ ; \ D_h D_k = 0 \ ( h \neq k )
\ ; \nonumber \\
D_h^{m_h} &=& 0 \ {\rm where} \ m_h = {\rm dim} P_h \ .
\label{A3}
\ea
In the case of Eq.~(\ref{35}), there are $s = 2 N$ distinct eigenvalues.
This is true as long as the positive and negative energy spectra do not
touch and is a consequence of level repulsion amongst the positive
eigenvalues and amongst the negative eigenvalues. Therefore there are
$2 N$ different orthogonal projectors $P_h$ which obey $\sum_h P_h = 1$.
This is possible only if ${\rm dim} P_h = 1$ for all $h = 1, 2, \ldots,
2 N$. But $m_h = 1$ for all $h$ implies $D_h = 0$ for all $h$ and
Eq.~(\ref{A1}) reduces to
\be
R(\zeta) = \sum_{h = 1}^{2 N} (\lambda_h - \zeta)^{-1} P_h \ .
\label{A5}
\ee
Since ${\rm dim} P_h = 1$ we have ${\rm Trace} P_h = 1$ for all $h$.
Taking the trace of Eq.~(\ref{A5}) and identifying the variable $\zeta$
with the energy, we obtain Eq.~(\ref{35}).

\end{document}